\documentclass[conference]{IEEEtran}
\IEEEoverridecommandlockouts
\usepackage{cite}
\usepackage{amsmath,amssymb,amsfonts}
\usepackage{algorithmic}
\usepackage{graphicx}
\usepackage{textcomp}
\usepackage{xcolor}
\usepackage{subfig}

\makeatletter
\newcommand{\linebreakand}{%
  \end{@IEEEauthorhalign}
  \hfill\mbox{}\par
  \mbox{}\hfill\begin{@IEEEauthorhalign}
}
\makeatother

\def\BibTeX{{\rm B\kern-.05em{\sc i\kern-.025em b}\kern-.08em
    T\kern-.1667em\lower.7ex\hbox{E}\kern-.125emX}}

\title{A Versatile Wireless Network Protocol for Spectrum Sharing with Passive Radio Services\\

}

\author{\IEEEauthorblockN{Ashton Palacios, Dinah Bronson, Jon Backman, Karl Warnick, Philip Lundrigan}
\IEEEauthorblockA{\textit{Department of Electrical and Computer Engineering} \\
\textit{Brigham Young University}\\
Provo, Utah, USA \\
apal6981@byu.edu, dinahsquire1@gmail.com, backmanjo908@gmail.com, warnick@ee.byu.edu, lundrigan@byu.edu}
}
\begin{document}
\maketitle

\begin{abstract}
    With the proliferation of wideband active services in bands shared with passive receivers for remote sensing and radio astronomy, new methods are needed for deconflicting active and passive users. We have developed a technique for active/passive user coordination that is compatible with essentially any existing wireless communications protocol. The passive user transmits an on-off keying modulated signal that can be detected by active radios using simple channel power estimates. Using off-the-shelf WiFi and LoRa hardware and on a software defined radio implementation of LTE, we show that Dynamic Passive to Active Spectrum Sharing (DPASS) is effective on a wide range of frequencies and physical layer implementations. We validate the protocol using these three technologies by demonstrating that each device receives a DPASS packet and dynamically takes an appropriate spectrum coordination action, including shutting off transmissions or switching frequencies.
\end{abstract}


\section{Introduction} \label{intro}
According to Cisco~\cite{cisco}, there will be over 29.3 billion network connected devices in use around the world by the end of 2023. Many of these devices are wireless that are able to communicate with devices of the same wireless protocol type while causing inter-technology interference with devices that do not use the same wireless protocol. Active devices are designed in such a way that mitigate this interference: IEEE 802.11 dictates a carrier sensing multiple access (CSMA) scheme, cellular providers maintain exclusive access to certain frequency bands, and LoRa uses chirp spread spectrum. While these and other solutions work for their respective protocols and devices, these solutions do not equate to spectrum coordination. Spectrum coordination allows heterogeneous devices to coordinate how and when to share a given radio frequency band. With more and more wireless devices coming online each year, \textit{explicit spectrum coordination is becoming paramount}.

\begin{figure}[b]
    \centering
    \includegraphics[width=.8\columnwidth]{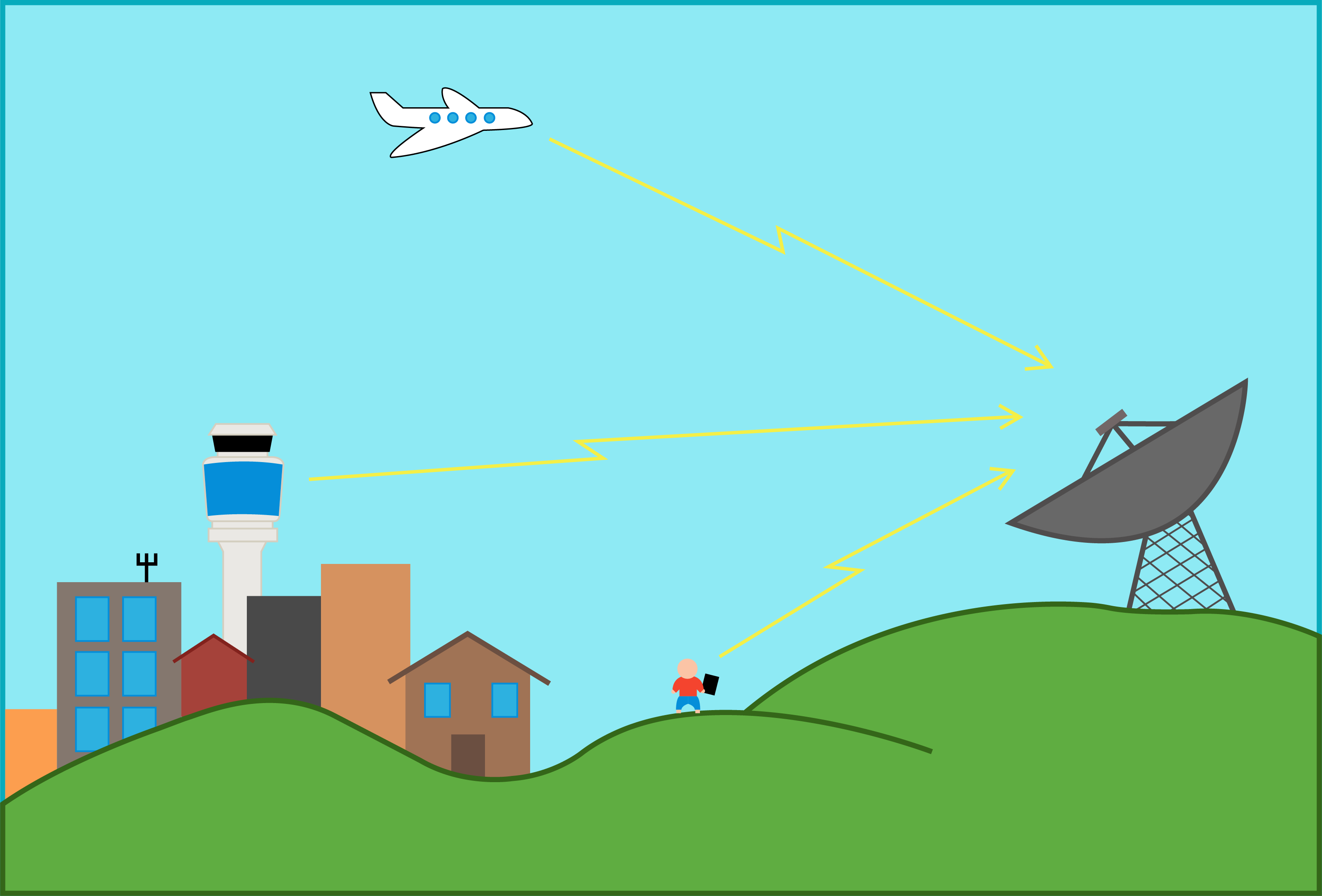}
    \caption{Active users interfering with a passive device, a radio astronomy telescope.}
    \label{fig:radio_telescope}
\end{figure}

Coordination with \textit{passive spectrum users} is becoming ever more important. An example of a passive spectrum user are astronomical receivers which are highly sensitive and detect signals many tens of dB below the thermal noise floor. Observation frequencies can range from 360~kHz to over 275~GHz~\cite{radio_regulations_2020} and the bandwidth can be anywhere between a few MHz to hundreds of MHz with observations lasting from a few minutes to hours. If there are any active transmissions from a wireless device within range of the telescope and on the frequency the telescope is listening to, the signals from space become nearly impossible to detect. The data at the time of interference often have to be ``blanked'' (i.e., deleted) because it is difficult to remove the interference and recover the signal. Having to blank a signal is a huge waste of time and money for scientists using radio observatories. Fig.~\ref{fig:radio_telescope} illustrates the many sources of frequency noise a radio astronomy telescope can encounter while trying to listen for cosmic electromagnetic radiation.

To mitigate interference, some passive observatories are located in radio quiet zones. These quiet zones have regulations that only allow certain transmitters and power levels at certain distances from the telescope~\cite{nat_quiet_zone}. The regulations block use of most electronic devices: cellular devices, microwave ovens, WiFi, and even gas powered vehicles, since the spark plugs emit electromagnetic interference. These restrictions are difficult to enforce. A rogue sensor interfering with an observatory could take valuable sensing time to track down and shut off, leading to a significant loss of data and ultimately money. Radio observatories along with other passive devices like aeronautical radio altimeters and passive sensing radar modules need a way to coordinate spectrum use with active transmitters.

Spectrum coordination between diverse wireless protocols is not easy. The physical layers of heterogeneous devices are often not compatible with each other. A Zigbee device using offset quadrature phase-shift keying modulation cannot directly communicate with a WiFi device using an orthogonal frequency-division multiplexing scheme even though they use the same frequencies. This leads to spectrum coordination solutions requiring a centralized controller, such as Citizens Broadband Radio Service (CBRS)~\cite{cbrs}. All participating devices in this scheme must be connect to a central CBRS controller called a spectrum access system (SAS). Before an active device transmits, it checks with the SAS to see if there are any limitations on the spectrum for a given time and location by uploading precise latitude and longitude coordinates. The CBRS service will inform the device if it can use the spectrum. If a dynamic change to the spectrum needs is required, the necessary data will not be published to CBRS connected devices in real-time. An additional limitation to using a system like this is that some devices do not have the capability to connect to this type of controller by either not having an internet connection or the ability to provide a precise location measurement. While a SAS works well for several types of devices and use cases, it is not a scalable solution for all devices, especially already deployed devices.

Recent research shows promise in the area of cross technology communication (CTC) as a viable technique to bridge the gap between heterogeneous device incompatibilities~\cite{CMorse, Esense, FreeBee, WEBee, WiBeacon, TamingCTC2021}. CTC is defined as exchanging data between devices that use heterogeneous wireless protocols. These techniques use two general methods: 1) emulating physical layer properties~\cite{TamingCTC2021, WEBee, WiBeacon} and 2) use of timing characteristics of transmissions \cite{CMorse, FreeBee, Esense}. Though the physical layer emulation techniques are able to achieve high data rates, they are a one-to-one solution: one physical layer is only able to emulate one other physical layer. This method exploits specific RF characteristics shared between the two target protocols, and as a result, these solutions do not work outside of their target protocols. CTC solutions using timing characteristics for communication only need a way of polling the wireless channel through received signal strength indicators (RSSI). These techniques have a slower data rate, but they have the capability to communicate to many types of heterogeneous devices, assuming they are on the same frequency band. However, implementations of these techniques often require specialized hardware, changes to firmware, or do not run in real time. They are also solely designed for active users.

For a spectrum sharing solution to be viable for \textit{passive users} it needs to have the following features:

\begin{itemize}
    \item \textbf{Dynamic}. It is not feasible for a device to have individual integrations with every wireless protocol, current and future. Radio astronomy exemplifies this fact; the potential frequency range of observations are between 360~kHz to over 275~GHz~\cite{radio_regulations_2020}. It is clearly not practical for a device to individually communicate with all possible wireless protocols within a given frequency band. A solution must be dynamic in nature, allowing a passive device to coordinate with any wireless device, regardless of wireless protocol.

    \item \textbf{Distributed}. As highlighted previously, spectrum coordination has been shown to work using a centralized server like CBRS. However, this requires consistent access to the server and advance scheduling of how the spectrum will be used. Our vision of spectrum sharing includes offline nodes and real-time dynamic usage of the spectrum. As a result, a distributed solution is required to accommodate all device types and usage patterns. In practice, a distributed method can be combined with the centralized approach.

    \item \textbf{Deployable}. With so many wireless devices already deployed, it is impractical for a solution to require changes to firmware or hardware. For a solution to be deployable and actionable now, active users should only need a software upgrade to become spectrum sharing participants.

\end{itemize}

To actively protect passive users, we develop a novel protocol, named the Dynamic Passive to Active Spectrum Sharing (DPASS) protocol. To the best of our knowledge, this is the first protocol designed for and protects passive users. The DPASS protocol allows us to create a versatile beacon that can be \textbf{received by any receiver}. Passive devices can use this new capability to notify nearby interfering active users to stop transmitting or switch to a different frequency. The DPASS protocol saves radio astronomy researchers valuable time and money and creates a framework for protecting other passive users. Our protocol is part of the vision of National Radio Dynamic Zones (NRDZ). For example, when a device enters a specific geographic area, our protocol would be enabled, allowing for spectrum coordination between passive users and active users.

\begin{figure}[b]
    \centering
    \includegraphics[width=\columnwidth]{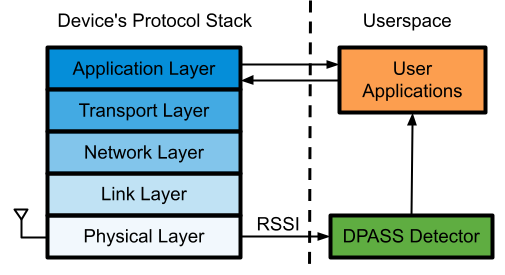}
    \caption{Generalized flow of a DPASS receiver. The receiver processes RSSI values in userspace that are provided from a lower protocol layer. When a DPASS packet is received, the DPASS detector notifies the user application to perform a spectrum sharing action.}
    \label{fig:process_flowgraph}
\end{figure}

Our protocol works by taking advantage of the lowest common denominator amongst receivers: the ability to measure the energy of the channel. By definition, all wireless receivers are capable of measuring the energy of a channel in one way or another. We exploit this ability to create a wireless protocol that all receivers can receive and decode that does not require any specialized receiving hardware. This allows devices to receive and decode our protocol from the application layer in userspace. Fig.~\ref{fig:process_flowgraph} shows a generalized flow for this process.

For this protocol to work, a passive user needs access to active transmitting hardware. The hardware needs to have the capability to transmit on the frequencies the passive user wants to listen to. It seems counterintuitive for a passive device to be combined with an active transmitter since the act of transmitting will cause interference with the passive device, which is what we are trying to mitigate. However, the transmitter hardware can be used in two ways that does not add additional interference. Before the passive device is listening, the transmitter can proactively notify nearby active devices that the passive device will be listening to a certain part of the spectrum soon. If the passive device is already listening and interference is occurring, the passive device transmitter can dynamically communicate with the interfering devices. Since interference is already occurring, the passive device transmitter is not adding any additional interference to the passive device. In the context of radio astronomy telescopes, a simple transmitter can be attached that coordinates with the telescope to broadcast its spectrum usage (frequency, bandwidth, and time) proactively and reactively. After receiving the DPASS packet, the active user can move to a non-interfering frequency or stop communication completely.

We demonstrate the dynamic nature of our protocol by showing its successful implementation with three diverse protocols: WiFi, LTE, and LoRa, which are widely used in billions of devices ranging from cellular phones to IoT sensors. We develop an application for each protocol that retrieves real-time RSSI estimations, decodes the DPASS protocol, and takes the necessary action to share the spectrum. Our system can work on frequencies beyond those used by these protocols, but these protocols can operate on frequencies that cover where many commercial devices operate. When a DPASS packet is received, the DPASS receiver decides on the appropriate spectrum sharing action, which varies depending on the targeted technology. For WiFi or LTE devices, the action is to switch to another RF channel outside the requested band, while for LoRa devices, it may need to stop transmitting entirely. We demonstrate that our protocol works under a variety of conditions to show its dynamic ability in communicating with a wide range of heterogeneous devices.


Our contributions in this paper are as follows:

\begin{enumerate}
    \item We introduce our novel protocol DPASS, which allows a transmitter to send a universal spectrum sharing beacon to \textit{heterogeneous receivers at once}. While we apply this technique to passive-to-active device coordination, our protocol could be used for many other spectrum sharing tasks.

    \item We design and build a DPASS transmitter on top of software defined radio (SDR) technology. We design a system that can modulate DPASS with over 300~MHz of bandwidth. Our solution has low overheading making it easy to integrate into existing SDR solutions. The transmitter code is publicly available on GitHub~\cite{github}.

    \item We design and integrate DPASS receivers into three technologies, WiFi and LoRa, using commercial off-the-shelf (COTS) hardware and without requiring any changes to the hardware, and srsRAN~\cite{srsran}. We evaluate that the three technologies are able to receive the DPASS protocol in real-time while the underlying technologies can still perform their normal functions. The receiver code is publicly available on GitHub~\cite{github}.
\end{enumerate}


\section{Design}
To be able to dynamically communicate with any wireless device, our design has to accommodate widely differing physical layer implementations. Unlike other solutions that create point-to-point integrations between different wireless protocols, our protocol works with all wireless devices \textit{simultaneously}. This makes our protocol highly scalable to the number of devices and number of wireless protocols. To do this, we focus on the fundamental capability of all receiving devices: the ability to measure the energy of a channel, often in terms of RSSI. We use a transmitter that sends out on-off keying (OOK) transmissions that can be measured by the receiver. These wireless environment changes can then be decoded by the device by correlating the signal strength measurements against a predetermined symbol. If a valid sequence of known symbols is detected, the active device interprets the symbols and reacts to the command. This can be done while the device is performing its normal operation. Because of the simplicity of the design and access to RSSI from software, our design can be implemented in software on virtually all wireless receivers.

\subsection{System Overview}
\begin{figure}[t]
    \centering
    \includegraphics[width=\columnwidth]{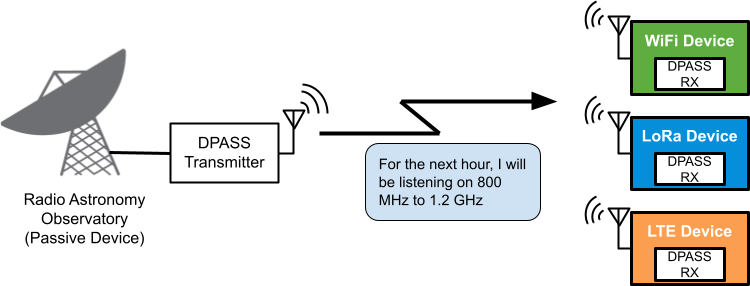}
    \caption{A high-level overview of our system. A passive user (e.g., a radio astronomy observatory) connects to a DPASS transmitter to relay observation times and frequencies. This information is transmitted to receivers on various device types running the DPASS software.}
    \label{fig:overview_img}
\end{figure}

The DPASS protocol consists of three entity types, as seen in Fig.~\ref{fig:overview_img}: the passive RF receiver that is trying to mitigate interference, the DPASS transmitter, and the active spectrum user(s). The DPASS transmitter and passive device can be one device, but for the rest of the paper we assume they are two entities. The system can work in two ways. Proactively, the passive user informs the active users through the transmitter beforehand of the duration of the listening cycle and the frequencies needed. Reactively, the passive user identifies interference and then informs the active users of the interference. This paper does not focus on how the passive user identifies interference, but rather on the system the passive user can use to mitigate interference. The DPASS transmitter receives the spectrum sharing information from the passive user and converts it to a DPASS packet. The DPASS transmitter then transmits the packet on the passive user specified frequencies that require coordination. The active users are constantly sampling the channel and detecting if a DPASS packet is present. Since many receivers measure RSSI for their normal operation, such as for clear channel assessment, this process adds little overhead to the receiving devices.

The following sections describe in detail how the passive user, DPASS transmitter, DPASS receiver, and DPASS protocol integrate to create a cohesive solution for spectrum sharing.

\subsection{Passive Device}
The passive device's primary purpose is to listen to the spectrum. The specifics of the passive device are outside the scope of this paper. Any device listening to the spectrum is susceptible to in- or out-of-band interference. Using DPASS, a passive device has two options---proactively or reactively mitigating the interference. Either way, the passive user needs to identify which frequencies need to be cleared and for how long. Once identified, this information is encoded into a DPASS packet to be transmitted out by a connected DPASS transmitter. If a transmitter is not already integrated with the passive device, one needs to be connected.

\subsection{DPASS Transmitter}
To be dynamic, a DPASS transmitter needs to be able to communicate to the lowest common denominator of receivers, the ability to measure the channel's energy level. Incidentally, any transmission has an effect on the energy level of the channel. A DPASS transmitter exploits this fact by using arguably the simplest modulation scheme, on-off keying (OOK), to manipulate the channel's energy level to encode data. To improve reliability of the transmissions, we use a technique similar to direct-sequence spread spectrum (DSSS). 

\subsection{DPASS Receiver}
Any device that has the ability to measure its channel's energy level can receive and decode a DPASS packet. By storing the most recent samples, we correlate channel energy samples against predetermined DPASS symbols. DPASS symbols are discussed in detail in Section~\ref{sec:DPASS_protocol}. Since the entire correlating process has to be completed each RSSI sample, a balance needs to be struck between the correlating algorithm speed and detection accuracy. For our implementation, we take advantage of some approximations to allow our program to run in linear time.

The detection/correlating algorithm has a few steps. It first receives and stores a new RSSI sample. For each sample it collects, it mean normalizes the collection of samples. The algorithm then correlates predetermined symbols against the normalized samples. If the correlation value is higher than a dynamically set threshold based on the variance of the collected samples, a symbol detection event is noted. If a valid sequence of symbols is detected within an appropriate time span, the DPASS packet symbols are decoded into a spectrum sharing action the active user should take.

\subsection{DPASS Protocol} \label{sec:DPASS_protocol}
The building blocks of a DPASS packet are sequences of pseudo-random number (PN) codes.  The DPASS protocol needs to be detectable amongst other transmissions. We make use of maximum length sequences (m-sequences) as our PN code sequence. These sequences have favorable cross-correlation properties~\cite{mls_corr}, making them a prime choice this application. M-sequences have varying lengths of powers of two minus one where individual bits within the sequence are called chips. Longer sequences have increased resistance to noise, but take longer to transmit, reducing throughput. We select a size of 63 chips as a balance between throughput and coding gain. There are 6 m-sequences with a size of 63 chips along with their inverses, totalling 12 usable symbols to design our protocol.

\begin{figure}[]
    \centering
    \includegraphics[width=.9\columnwidth]{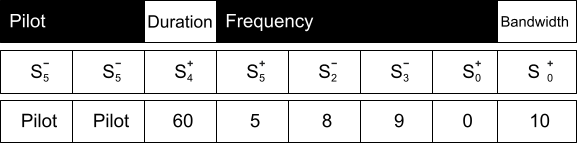}
    \caption{The DPASS packet structure (top), an example packet with its corresponding symbol sequence (middle), and the meaning of the example packet (bottom).}
    \label{fig:protocol_design}
\end{figure}

We design the DPASS protocol message structure to use as little air time as possible. By doing so, we minimize both interference and time of detection. The DPASS protocol structure consists of a sequence of eight symbols. Table~\ref{tab:symbol_definition} provides a comprehensive list of the different meanings of each symbol under the different symbol locations within a DPASS packet. Fig.~\ref{fig:protocol_design} shows an example packet with its corresponding symbols and meanings. The packet is broken into a preamble and a payload. The first two symbols constitute the preamble which can be synchronized against. The active user can use the preamble to identify that a DPASS payload is about to be received. The first symbol in the payload is an action time duration symbol. When an active user receives and decodes a DPASS packet, an appropriate spectrum sharing action needs to be performed for a certain amount of time. Radio astronomy telescopes listen to the spectrum from a few minutes to a few hours~\cite{dynamic_scheduling_system}, so action times are to be interpreted in minutes. Following the duration symbol, the next four symbols constitute a center frequency in MHz that the passive user is requesting. This gives the protocol an accuracy of one MHz. The final symbol to be transmitted informs the active user of the bandwidth size the passive user is requesting. Many active devices have multiple channels that are continuous in frequency. This final symbol informs the receiving device whether or not the passive device also needs the surrounding channels to be free. To allow 12 symbols to express all of this information, the symbol location within the protocol gives a different meaning to the symbol. We have selected these values to fit the needs of radio astronomy. However, other values can be used for different applications.

For example, if a radio astronomy telescope wants to listen to the frequencies between 5.84 and 5.93~GHz for 50 minutes, it will create a DPASS packet with a duration of 60 minutes, a center frequency of 5.890~GHz, and a bandwidth of 10~MHz, as shown in Fig.~\ref{fig:protocol_design}. Compared to a solution like CBRS, DPASS requires no connection to a centralized server. All communication is done locally and in a distributed fashion.

\begin{table}[t]
    \centering
    \caption{The meanings of the symbols under the different protocol contexts. The subscript denotes the symbol number and the superscript denotes whether the symbol is inverted or not.}
    \renewcommand{\arraystretch}{1.2}
    \label{tab:symbol_definition}
    \begin{tabular}{rrrr}
        \textbf{}       & \textbf{Duration} & \textbf{}          & \textbf{Bandwidth} \\
        \textbf{Symbol} & \textbf{(Min)}    & \textbf{Frequency} & \textbf{(MHz)}     \\ \hline
        $S_0^+$         & 5                 & 0                  & 10                 \\
        $S_1^+$         & 10                & 1                  & 20                 \\
        $S_2^+$         & 20                & 2                  & 40                 \\
        $S_3^+$         & 40                & 3                  & 80                 \\
        $S_4^+$         & 60                & 4                  & 160                \\
        $S_5^+$         & 90                & 5                  & 320                \\
        $S_0^-$         & 120               & 6                  & 640                \\
        $S_1^-$         & 180               & 7                  & -                  \\
        $S_2^-$         & 240               & 8                  & -                  \\
        $S_3^-$         & 300               & 9                  & -                  \\
        $S_4^-$         & 360               & 10                 & -                  \\
        $S_5^-$         & PILOT             & PILOT              & PILOT
    \end{tabular}
\end{table}

\section{Implementation}
We discuss the implementation of DPASS on WiFi, LoRa, LTE devices, and a DPASS transmitter. We first develop a transmitter that can accommodate a wide range of spectrum coordination scenarios on an SDR. We use COTS hardware for WiFi and LoRa DPASS receiving devices. We continue to use SDRs for LTE, which is discussed in its respective subsection.

\subsection{Transmitter}
Our transmitter needs to be versatile and easy to use over a wide range of frequencies; radio astronomy telescopes operate in subsets of frequencies beginning at 360~kHz to over 275~GHz~\cite{radio_regulations_2020}. Since most consumer electronics can operate in the range between 10~MHz and 6~GHz, we opt to focus on this range, although the techniques we develop will work on other frequency ranges. We select the Ettus USPR X410 to be our transmitter. This radio supports an approximate frequency range of 10-7000~MHz. We choose to use an SDR because of its frequency range capabilities along with the fact that they are easy to configure and use. Our implementation uses GNU Radio~\cite{gnu} to create and modulate our DPASS packets. Using GNU Radio and the RFNoC capabilities of the X410, we create a modulated signal that utilizes over 300~MHz of instantaneous bandwidth, allowing our protocol to communicate with wireless devices that are spread over 300~MHz of bandwidth \textbf{at once}. This demonstrates that the transmitter can be flexible and powerful enough to accommodate many spectrum coordination scenarios. Our protocol's maximum bandwidth is limited by the bandwidth of the software defined radio (SDR).

\subsection{WiFi Receiver}

The WiFi DPASS receiver is built upon COTS hardware without needing to modify any hardware or firmware. We select WiFi NIC with the Atheros ATH9K chipset. The computers in which these are installed use Ubuntu 20.04. The ATH9k chipset has a specialized mode~\cite{spectral_scan} that provides physical and MAC layer information, like RSSI, to the Linux virtual debug file system in userspace. The passing of this information from kernel space to userspace is shown in Fig.~\ref{fig:process_flowgraph}. The mode also provides a few control files that define the underlying information reporting behavior. These files allow us to control how often new information is provided to userspace. We use a sample rate that allows us to detect 5ms symbol chips. We use an application layer Python script that configures, reads and processes samples, and then takes spectrum sharing actions upon DPASS packet detection from these debug files.

Our WiFi implementation has two spectrum sharing options. It can either shut off the entire radio or switch to another frequency band. We use the Linux command line program \texttt{nmcli}~\cite{nmcli} to control different functionalities of the WiFi adapter, such as turning the radio off or switching to different frequency bands.

\subsection{LoRa Receiver}

We implement a LoRa DPASS receiver on the COTS Pycom LoPy4 board~\cite{pycom}. The LoPy4 board is a development board that comes with LoRa, Sigfox, WiFi, and BLE capabilities that can be programmed with MicroPython.

To make the underlying RSSI values available to the userspace software in MicroPython, we make a minor addition to the default Pycom Python bindings. Our program can reliably sample the channel's energy with intervals as low as 500 $\mu$s. For consistency and in order to avoid excess load on the processor, we sample at 5 ms intervals. Because of hardware limitations, the LoPy4 board cannot report valid RSSI values during active transmissions. However, because of the LoRaWAN 1\% duty cycle~\cite{lora_parameters} for many devices, this only minimally affects the device's response for most realistic scenarios.

To ease implementation complexity, much of the data processing for the LoRa receiver is done on an external computer running a similar Python script that the WiFi implementation uses while the LoPy4 board is connected and reporting RSSI through serial communication. The board sends the current measured RSSI values which are processed in real-time. Unlike the WiFi implementation, our LoRa receiver does not have another frequency it could switch to upon receiving the DPASS packet. Instead, after receiving a DPASS packet, the device is put to sleep for the specified amount of time.

\subsection{LTE Receiver}

Because of government regulations for transmitting on cellular frequencies, our LTE implementation uses srsRAN, a 4G/5G SDR implementation of the LTE protocol stack~\cite{srsran}. The use of an SDR enables us to transmit on non regulated ISM bands instead of cellular bands. Additionally, srsRAN provides a lot of real time information about the physical layer. However, the default provided RSSI value for srsRAN is only reported in conjunction with a valid packet reception. To access the signal strength readings regardless of packet reception, we modify the srsRAN source code. We access the received I/Q input buffer and calculate an RSSI value once every ms. We then expose this value through a ZeroMQ (ZMQ)~\cite{zeromq} socket.

We use a pair of Ettus USRP B200 minis to simulate a user equipment (UE) device and an evolved Node B (eNodeB). To ensure connectivity, \texttt{iperf3}, a command line network throughput evaluation application, is used between the UE to the eNodeB. While srsRAN takes care of the active connection, a userspace program reads and processes the RSSI values from the ZMQ socket looking for DPASS packets. The two Ettus B200 minis run on separate computers and connect wirelessly in the 2.4~GHz ISM band. When a DPASS packet is received by the UE device, the UE device turns off. Consumer LTE devices can avoid turning off their radios by simply switch frequencies or enabling airplane mode.

\section{Evaluation}
We evaluate the DPASS characteristics and our active user implementations to show that DPASS packets can be received and decoded across different protocols from userspace. We demonstrate that DPASS packet spectrum bandwidth can have a varying range of bandwidth. We also show how DPASS and native protocol transmissions affect each other. Finally, we show that reception of DPASS packets can elicit spectrum sharing actions from devices.

In terms of setup, the LoRa implementation uses the 915~MHz frequency band, WiFi uses channels in both 2.4~GHz and 5~GHz frequency bands, and our srsRAN LTE implementation uses the 2.4~GHz ISM band to avoid transmitting on licensed spectrum. For the experiments that use WiFi devices, we use the Linksys MR7500 MESH WIFI 6E wireless router as our access point. It is to be noted that there are many other devices outside the control of our experiments communicating on the 915~MHz, 2.4~GHz and 5~GHz bands we use in our experiments that introduce random interference.

\begin{figure}[t]
    \centering
    \includegraphics[width=.9\columnwidth]{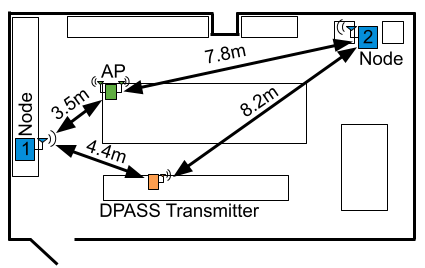}
    \caption{Overview of the placement of the DPASS enabled computers, access point, and the DPASS transmitter in our lab space. Other non-labeled boxes are desks, arm chairs, and other office furniture items.}
    \label{fig:wifi}
\end{figure}

\begin{table}[tb]
    \caption{Short-range proof of concept results of WiFi, LoRa, and LTE devices receiving DPASS packets.}
    \label{tab:proof_of_concept}
    \resizebox{\columnwidth}{!}{%
        \begin{tabular}{rrrrr}
            \textbf{}       & \textbf{DPASS} & \textbf{DPASS} & \textbf{False}     & \textbf{False}     \\
            \textbf{Device} & \textbf{TX}    & \textbf{RX}    & \textbf{Positives} & \textbf{Negatives} \\ \hline
            WiFi 1          & 104            & 104            & 0                  & 0                  \\
            WiFi 2          & 104            & 104            & 1                  & 0                  \\
            LoRa            & 160            & 160            & 0                  & 0                  \\
            LTE             & 104            & 104            & 0                  & 0                  \\
        \end{tabular}
    }
\end{table}

\subsection{Symbol Reception}
To ensure DPASS works with our WiFi, LoRa, and LTE implementations, we perform symbol and DPASS packet reception tests in which we transmit out a known number of packets and evaluate how many are detected and properly decoded. For WiFi, we use the physical configuration depicted in Fig.~\ref{fig:wifi} to mimic real world placement WiFi enabled desktop computers and to show that multiple devices can receive the DPASS packet at the same time. We perform the LoRa experiment outside with clear line of sight with 357 meters between the DPASS transmitter and the LoRa device. LTE is performed with two meters between the DPASS transmitter and LTE receiver. The results are shown in Table~\ref{tab:proof_of_concept}. The results show that each of these devices can receive and detect DPASS packets and the underlying symbols. This shows that DPASS packets can be detected at range and in scenarios that mimic real life.

\subsection{Transmitter Capability}
\begin{figure}[t]
    \centering
    \includegraphics[width=\columnwidth]{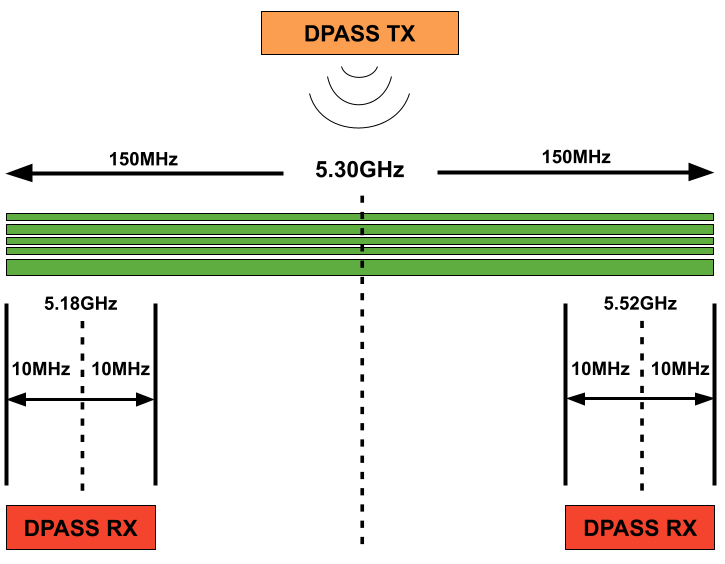}
    \caption{DPASS transmitter generates a DPASS packet with over 300~MHz of bandwidth that is detected and decoded by WiFi devices at the edges of the signal.}
    \label{fig:wideband}
\end{figure}

Passive devices have widely varying spectrum needs, and a DPASS transmitter needs to be designed to accommodate these needs. We first configure our transmitter to target LoRa with a small bandwidth of 2~MHz. Next, we configure our transmitter to generate a packet with over 300~MHz of bandwidth that targets WiFi or LTE devices on vastly different frequencies. To test out the DPASS transmitter, we broadcast on 5.3~GHz using a USRP x410, while two independent WiFi receivers are on 5.18~GHz and 5.52~GHz respectively, Fig.~\ref{fig:wideband}. The two receivers are able to decode the packet at the same time. This experiment shows that the transmitter is capable of transmitting narrow and wide band packets. This allows a DPASS transmitter to fit the specific needs of the passive device whether that is mitigating interference a single type of device on a small band of frequencies to many different types of devices on a large band of frequency. The limiting factor of how wide the transmission can be scaled is limited by the transmitting hardware and not our protocol.

\subsection{Network Impact}\label{sec:Network_Utilization}
With normal device communication and DPASS packet transmissions potentially coinciding in time and frequency, it is easy to ask how DPASS affects the primary channel of communication and how the primary channel affects DPASS. To examine this duality, we use WiFi nodes 1 and 2 from Fig.~\ref{fig:wifi}. Using \texttt{iperf3}, we create a client-server connection between nodes 1 and 2 to simulate network traffic in conjunction with interference from other wireless devices already communicating on the same frequency. Node 1 is a wired server and node 2 is the wireless client. The client connects to the server via the wireless router and sends UDP packets at different throughput values. Only node 2 in this configuration is running the DPASS detection algorithm. The spectrum sharing action feature is disabled for this test to allow capturing multiple DPASS packets in a single test. We run two sets of tests: a benchmark test where no DPASS packets are transmitted and one where 10 DPASS packets are transmitted. Each test is repeated 5 times for each of the selected throughput rates, which are unlimited, 25~Mbps, and 10~Mbps.

\begin{figure}[t]
    \centering
    \includegraphics[width=\columnwidth]{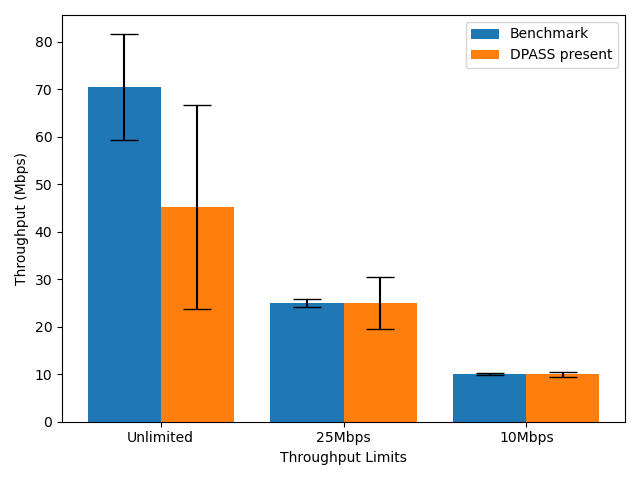}
    \caption{Average throughput with and without DPASS packets being transmitted.}
    \label{fig:net_util}
\end{figure}

\subsubsection{Affect of Network Traffic on DPASS} The DPASS modified WiFi node is able to fully decode all but one out of the 150 DPASS packets transmitted under all the different network traffic configurations.  This demonstrates that DPASS works despite the amount of network utilization from other active wireless transmissions. The use of DSSS in the protocol's design allows OOK to be detected regardless of other transmissions. In addition, the DPASS transmitter does not adhere to any channel sensing scheme and instead transmits whether or not the channel is busy. This shows that our design will work even if the channel is congested.

\subsubsection{Affect of DPASS on Network Traffic} The DPASS protocol uses active transmission, which will affect the throughput of other WiFi transmissions. Fig.~\ref{fig:net_util} shows the impact of the DPASS protocol on other network traffic. The unlimited throughput configuration shows an average drop in throughput of 36\% when DPASS is running compared to when it is not. However, these results show expected behavior. WiFi's CSMA mechanism prevents WiFi transmissions while the channel is busy. DPASS transmissions put energy into the spectrum that WiFi devices detect as interference and will not transmit at the time, leading to lower throughput values as shown by the unlimited throughput configuration. DPASS has little to no average effect when the traffic limit is set to 25~Mbps and 10~Mbps, which are more realistic scenarios. These scenarios have a worst case drop of roughly 20\% of throughput. Dynamic spectrum sharing (DSS) is a physical layer spectrum coordination technique between 4G and 5G cellular devices that is becoming widely adopted. It is reported to have a network performance impact to all users of up to 25\%, whether or not these devices adhere to the DSS system~\cite{dss}. The DPASS protocol shows similar results in network degradation demonstrating that the worst case scenario is within an acceptable range while there is little to no impact under normal usage conditions. This is important because there may be higher priority communications that will need to interrupt passive users regardless, like emergency communications. These communications, under normal throughput usage, will experience no impact and ignore the DPASS protocol.







\begin{figure*}[ht]
    \centering
    \includegraphics[width=\textwidth]{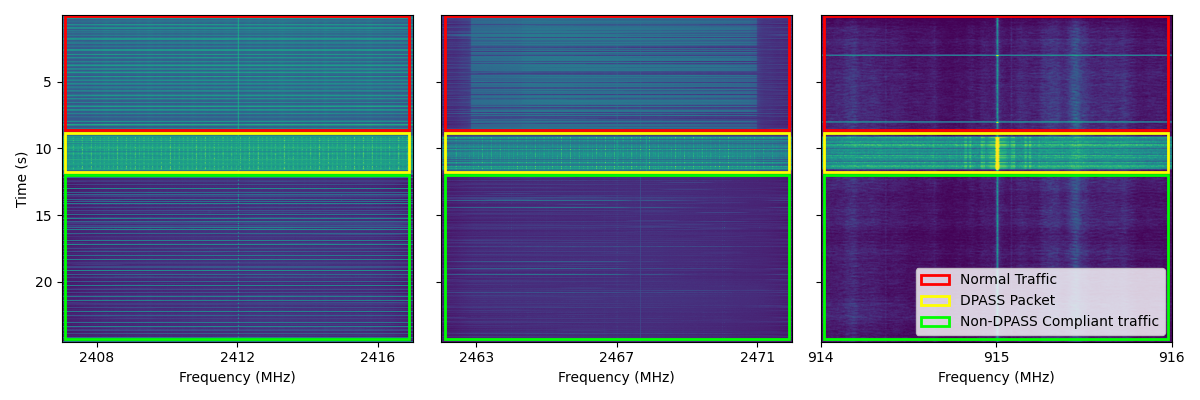}
    \caption{Spectrograms showcasing three distinct wireless protocols receiving and acting on a DPASS packet. Left: WiFi, Middle: srsRAN, Right: LoRa. The top portion encircled in red is the normal device communication. The yellow box is the transmitted DPASS packet. The bottom green portion is other non-DPASS compliant traffic that was on the same frequency.}
    \label{spectrogram}
\end{figure*}

\subsection{Multi-Protocol Coordination} 
We implement the detection of the proposed protocol on three vastly different protocols: WiFi, LoRa, and LTE. In this experiment, we show that we can coordinate with all three device types from one transmitter using one protocol, DPASS. For simplicity of setup we place our LoRa, WiFi, and srsRAN devices on a desk approximately a meter from our DPASS transmitter. The LoRa device periodically transmits data every 5 seconds to another LoRa device across the room. The WiFi device is connected to our access point which operates on channel 1 (2.412~GHz center frequency). The uplink frequency for the srsRAN devices is set to the center frequency of WiFi's channel 12 (2.467~GHz). The WiFi and srsRAN devices use \texttt{iperf3} to simulate network traffic. The spectrogram plots found in Fig.~\ref{spectrogram} show the normal spectrum use of the individual devices between 0 and 9.5 seconds as denoted by the red boxes in the figures. It is to be noted that the LoRa sub-figure does not exhibit the tell-tale chirp-spread spectrum pattern that is normally associated with LoRa because of the time scale in the figure. At 9.5 half seconds a DPASS packet is transmitted. For Lora, the packet is transmitted on 915~MHz using our 2~MHz bandwidth transmitter, while for the other devices the DPASS packet is transmitted on channel 6 (2.437~GHz center frequency) using our wideband transmitter. After the devices receive and decode the DPASS packet, they take appropriate spectrum sharing actions. The LoRa and srsRAN devices merely shutoff their radios for the rest of the experiment. The WiFi device, on the other hand, switches to the access point's 5~GHz interface and resumes transmitting its normal network traffic. From this experiment, it is evident that the DPASS protocol can communicate with many different devices and coordinate the use of spectrum. The DPASS protocol is dynamic in its bandwidth usage and the devices it can be received on. Also, in comparison to the CBRS approach, DPASS does not require the complexity of contacting a SAS to check if spectrum use is allowed. The proposed solution does not need SAS enabled hardware or changes to the current hardware to receive and decode DPASS messages.

\section{Related Work}

\subsubsection*{Cross-Technology Communication Protocols}
Several CTC \\methods have been introduced in recent literature. Some of them involve manipulating the transmission from one device to emulate a signal that can be received by the other device of a different protocol. For example, others have created physical layer emulation techniques that allow a WiFi device to communicate with BLE devices ~\cite{TamingCTC2021, WiBeacon} or WiFi devices to Zigbee devices~\cite{WEBee}. These techniques are restricted to communication between the two protocols, making large-scale inter-technology communication systems difficult. Their focus tends to be on making CTC as fast as possible. The goal of our work is to provide a protocol that any receiver can use. We focus on the fundamental capabilities of general receivers at the expense of data rate. Our method does not use any of these emulation techniques.

Other CTC methods modify the timing characteristics of transmissions, such as beacons or data frames, to relay information from devices using one protocol to those using another. For example, FreeBee~\cite{FreeBee} shifts the timing of WiFi beacon frames to convey information, and C-Morse~\cite{CMorse} slightly adjusts data frame timing as data frames are sent out to provide communication. Most similar to our approach is Esense~\cite{Esense}, which uses received signal strength to communicate between otherwise incompatible physical layers. However, Esense uses the duration of the transmission to encode data, where our solution uses DSSS techniques. As a consequence, the Esense protocol can't be detected at range or in the presence of a moderate amount of noise. In summary, these methods are too complex to work over a wide range of devices. The DPASS system talks to the lowest common denominator of receiving devices in a simple manner, making it \textbf{dynamic}.

Other protocols have been developed that use changes in signal strength (or RSSI) to communicate with devices. However, these protocols focus on vastly different use cases, such as long distance communication~\cite{ONPC} or ultra-low-power wireless communication~\cite{wifi-backscattering}. Our protocol focuses on providing coordination between passive and active devices, and as a result the design of our protocol is unique in terms of overall architecture and implementation.

\subsubsection*{Spectrum Management Techniques}
A primary proposed alternative for regulatory spectrum management is having a centralized network such as the one proposed by \cite{ZyliniumResearch}, the coordination network of the CBRS bands~\cite{cbrs}, or the 5G DSS. While these techniques have proven to be effective, they have limitations. Mobile transmitters may not be aware of current zone restrictions~\cite{nat_quiet_zone} or they wander into a telescope's interference range. These cases cannot always be handled by a centralized network. Also, many low-power devices may not be capable of accessing said networks. In such cases, our DPASS solution would allow the passive user to do its own spectrum coordination with any device. In practice, both methods could be used to provide the best spectrum sharing protection for passive devices.

\section{Future Work} \label{sec:Discussion}

\subsubsection*{Security}
The ability to shut off arbitrary transmitters on a given frequency band is a powerful capability that could easily be misused. This paper serves as a proof of concept, so a complete discussion on how to mitigate the misuse of DPASS is beyond the scope of this paper. However, we briefly present some ideas on how we could achieve this. First, similar to how there are restrictions on what frequencies a transmitter uses, regulators could choose to make transmitting a DPASS signal on any frequency against the law unless they are an authorized node, such as a radio astronomy observatory. A DPASS transmission would be especially detectable (compared to other types of spectrum offenses) since it is broadcast in nature, and all nodes that receive the message could alert the authorities of unauthorized use. Second, the DPASS system could be adapted to include a secure token that is transmitted along with the spectrum usage information. A receiving node would follow the DPASS command only if the token is verified. We leave it to future work for an exploration of these different options.

\subsubsection*{Re-broadcast of DPASS Packets}
National quiet zones can include tens to hundreds of square kilometers of space \cite{nat_quiet_zone}. It would be infeasible for a transmitter to have the transmit power to cover that entire range. Devices that receive a DPASS packet could relay the spectrum sharing message through its native protocol to neighboring devices or gateways.  To protect against a flooding of these extended DPASS packets, the native protocol implementation of a DPASS packet could include a time-to-live header field.

Also, many IoT sensors go to sleep in between transmissions, causing sensor ``deafness'' where the sensors are not actively polling the channel to conserve battery. To overcome the device's deafness, the device's gateway would be the target of the DPASS packet. While some gateway schemes don't return acknowledgements back to their clients, many do in which the gateway could inform the sensor of the DPASS packet.


\section{Conclusion}
We propose and demonstrate the effectiveness of using explicit communication for coordination between passive and active users using the Dynamic Passive to Active Spectrum Sharing protocol. DPASS is a distributed solution that can accommodate the needs of devices that cannot be connected to a centralized spectrum coordination network, and it can be implemented in any type of wireless communication device from userspace. We show that WiFi, LoRa, and LTE devices can receive DPASS packets and make appropriate spectrum sharing actions. A DPASS packet can be received even when there are other active transmissions present. Combining this solution with other systems can further expand and increase the possibility of effective dynamic spectrum coordination.

\bibliographystyle{IEEEtran}
\bibliography{refs}


\end{document}